\documentclass[conference]{IEEEtran}
\usepackage{amsmath}
\usepackage{amssymb}
\usepackage{latexsym}
\usepackage{multirow}
\usepackage{epsfig}
\usepackage{graphics}
\usepackage{mathrsfs}
\usepackage{algorithm}
\usepackage{algpseudocode}
\usepackage{amsthm}
\usepackage{url}
\usepackage{color}
\usepackage{graphicx}
\usepackage{algcompatible}
\usepackage{array}
\usepackage{amsfonts}
\usepackage{mdwmath}
\usepackage{mdwtab}
\usepackage{eqparbox}
\usepackage{epstopdf}
\usepackage{cite}

\newcommand{\bc}{\begin{center}}
\newcommand{\ec}{\end{center}}
\newcommand{\be}{\begin{equation}}
\newcommand{\ee}{\end{equation}}
\newcommand{\bea}{\begin{eqnarray}}
\newcommand{\eea}{\end{eqnarray}}

\IEEEoverridecommandlockouts
\begin{document}
\title{WiSegRT: Dataset for Site-Specific Indoor Radio Propagation Modeling with 3D Segmentation and Differentiable Ray-Tracing
\\ \LARGE \emph{(Invited Paper)}}
\author{\IEEEauthorblockN{Lihao Zhang\IEEEauthorrefmark{1}, Haijian 
Sun\IEEEauthorrefmark{1}\thanks{The work of H. Sun was supported in part by NSF CNS-2236449, and The University of Georgia E-Mobility Initiative seed grant. The work of R. Q. Hu was supported in part by NSF CNS-2007995, ECCS-2139508, IMR-2319487.}, Jin Sun\IEEEauthorrefmark{2}, and Rose Qingyang Hu\IEEEauthorrefmark{3}} 
\IEEEauthorblockA{
\IEEEauthorrefmark{1}School of Electrical and Computer Engineering, University of Georgia, Athens, GA, USA \\
\IEEEauthorrefmark{2}School of Computing, University of Georgia, Athens, GA, USA \\
\IEEEauthorrefmark{3}Department of Electrical and Computer Engineering, Utah State University, Logan, UT, USA \\
Emails: \{lihao.zhang, hsun, jinsun\}@uga.edu, rose.hu@usu.edu}
}

\maketitle

\IEEEpeerreviewmaketitle
\begin{abstract}
The accurate modeling of indoor radio propagation is crucial for localization, monitoring, and device coordination, yet remains a formidable challenge, due to the complex nature of indoor environments where radio can propagate along hundreds of paths. These paths are resulted from the room layout, furniture, appliances and even small objects like a glass cup. They are also  influenced by the object material and surface roughness. Advanced machine learning (ML) techniques have the potential to take such non-linear and hard-to-model factors into consideration. However, extensive and fine-grained datasets are urgently required.  This paper presents WiSegRT\footnote{\url{https://github.com/SunLab-UGA/WiSegRT}}, an open-source dataset for indoor radio propagation modeling. Generated by a differentiable ray tracer within the segmented 3-dimensional (3D) indoor environments, WiSegRT provides site-specific channel impulse responses for each grid point relative to the given transmitter location. We expect WiSegRT to support a wide-range of applications, such as ML-based channel prediction, accurate indoor localization, radio-based object detection, wireless digital twin, and more. 

\end{abstract}


\section{Introduction}
\par The pursuit of advancements in wireless communication systems has led to the exploration of innovative technologies such as integrated sensing and communication (ISAC) \cite{ISAC_liu2022integrated}, reconfigurable intelligent surfaces (RIS) \cite{RIS_liu2021reconfigurable}, and massive multiple-input multiple-output (mMIMO) systems \cite{mimo_lu2014overview}. These technologies are  expected not only to enhance communication data rates but also to perform additional tasks such as  sensing and localization. The key to their success lies in the availability of accurate and comprehensive wireless channel information. However, obtaining this information poses a significant challenge in the field of radio propagation modeling. 

\par In general, radio propagation models are categorized into two types: probabilistic and deterministic models. Probabilistic models are grounded in empirical formulas and utilize statistical data to calibrate parameters for specific environments. These models, primarily based on the distance between the transmitter (Tx) and the receiver (Rx), are good at rapid signal strength estimation. However, They often lack accuracy and the ability to provide detailed channel characteristics such as the channel impulse response (CIR), angle of arrival (AoA), and angle of departure (AoD).

\par Deterministic models, in contrast, consider the precise environment features. Some deterministic models are simplistic, such as the free space or the two-ray ground reflection model. Some others employ computational electromagnetic (CEM) techniques \cite{CEM_bondeson2012computational}, which utilize Maxwell's equations with appropriate boundary conditions, making them suitable for small-scale and near-field analysis. For radio propagation in large-scale environments and the transmitter's far-field, ray concept from geometrical optics \cite{GO_born2013principles} is often employed. In a homogeneous medium, typically air, rays travel in straight lines, carrying energy and obeying the laws of reflection, refraction, and diffraction upon interacting with objects. This ray concept is the basis for implementing the well-known ray tracing method for both graphic rendering and radio simulation. However, both CEM methods and approximate ray tracing are computationally intensive and time-consuming, which limits their feasibility for real-time applications that also require high-quality results.

\par To address the issue of high computational burden associated with CEM and ray tracing, the progress in machine learning (ML) has given rise to the ML-based radio propagation models that offer remarkable capabilities and fast computation speed \cite{2d_bakirtzis2022deepray,2d_radio_levie2021radiounet,2d_seretis2022toward,3d_kocevska2023identification,winert_orekondy2022winert,nerf2,mmsv_kamari2023mmsv}. The effectiveness of neural network depends significantly on both the quantity and quality of the data. Although there are many radio propagation datasets for outdoor or indoor scenarios, such as\cite{dataset_viwi_alrabeiah2020viwi,dataset_zhang2021generalized}, the majority merely employ 2D layouts or primitive 3D shapes as the environment input and coarse channel characterizations, such as the received signal strength indicator (RSSI). This setting may suffice for outdoor scenarios, but often underperform in capturing the complexity of indoor environments, as will be discussed in a later section.

\par In this paper, we introduce \underline{Wi}reless \underline{Seg}mented \underline{R}ay \underline{T}racing (WiSegRT), a precise indoor radio dataset designed for various ML tasks in channel modeling. The main features and contributions are as follows:

\begin{itemize}

\item WiSegRT comprises 10 (more to come) realistic and high-definition indoor scenes with furniture, appliances, and decorations. Objects within these scenes possess distinct and detailed material segmentation, such as the glass panel, plastic casing, and metal stand of a television.

\item Radio simulation was conducted in \emph{Sionna} \cite{sionna_hoydis2022sionna}, a state-of-the-art, open-source, GPU-accelerated physical layer communication system simulator developed by NVIDIA. All of its components are differentiable and can be concatenated into neural networks for end-to-end tasks. 

\item The 3D scenes were created using \emph{Blender} software and support high-quality rendering since our scenes also have realistic rendering textures. Researchers can generate corresponding visual-based datasets to implement vision-based ML methods.

\end{itemize}

\section{Motivation and Dataset Overview}
\par  In this section, we introduce the motivation that have inspired us to construct this dataset, and also provide a brief overview of the dataset structure. 

\subsection{Motivation}
\par Recent advancements have increasingly leveraged ML approaches in radio propagation modeling tasks. Works such as  \cite{2d_radio_levie2021radiounet,2d_bakirtzis2022deepray, 
2d_seretis2022toward}, employed 2D environmental layouts as inputs to train neural networks for  radiomaps prediction, which represent signal strength across all the locations on the map. Some works, like \cite{3d_kocevska2023identification}, take simple 3D environments as inputs and train the networks to output radiomaps or more detailed channel characterizations like CIR, AoA, and AoD. In \cite{winert_orekondy2022winert}, the proposed \emph{WiNeRT} uses neural networks to replace the radio-object interactions. In \cite{nerf2}, the proposed \emph{NeRF$^2$} utilizes the idea of implicit representation from the neural radiance field and can output the spatial spectrum for any Tx-Rx pair in the trained scene. In \cite{mmsv_kamari2023mmsv}, the proposed method \emph{mmSV} uses the street view pictures as input, then reconstructs the 3D environment with the material assignment which are also from the street view pictures by semantic segmentation. There is a growing trend in ML-based radio propagation modeling, along with a substantial demand for high-quality training data.

\par However, existing channel datasets often neglect intricate details of indoor environments, such as potential reflection faces, scattering areas, and diffraction edges. Therefore, the model designed for naive 2D or 3D environment and trained on simple data may fail to grasp these rich features. Furthermore, the precision of material segmentation is crucial and neglecting it can result in unexpected high-reflective paths and introduce instability into the system. For more delicate ML-based wireless applications, like mMIMO and radio sensing, the accuracy of radio propagation modeling is the key. Thus, we build WiSegRT to provide a virtual reality for radio propagation. Thus researchers can not only obtain comprehensive radio propagation data from the elaborately created environment, but also generate corresponding visual data for vision-aided radio systems like in \cite{mmsv_kamari2023mmsv}. 

\subsection{Dataset Overview}

\begin{figure}[h]
    \centering
    \includegraphics[width=0.7\linewidth]{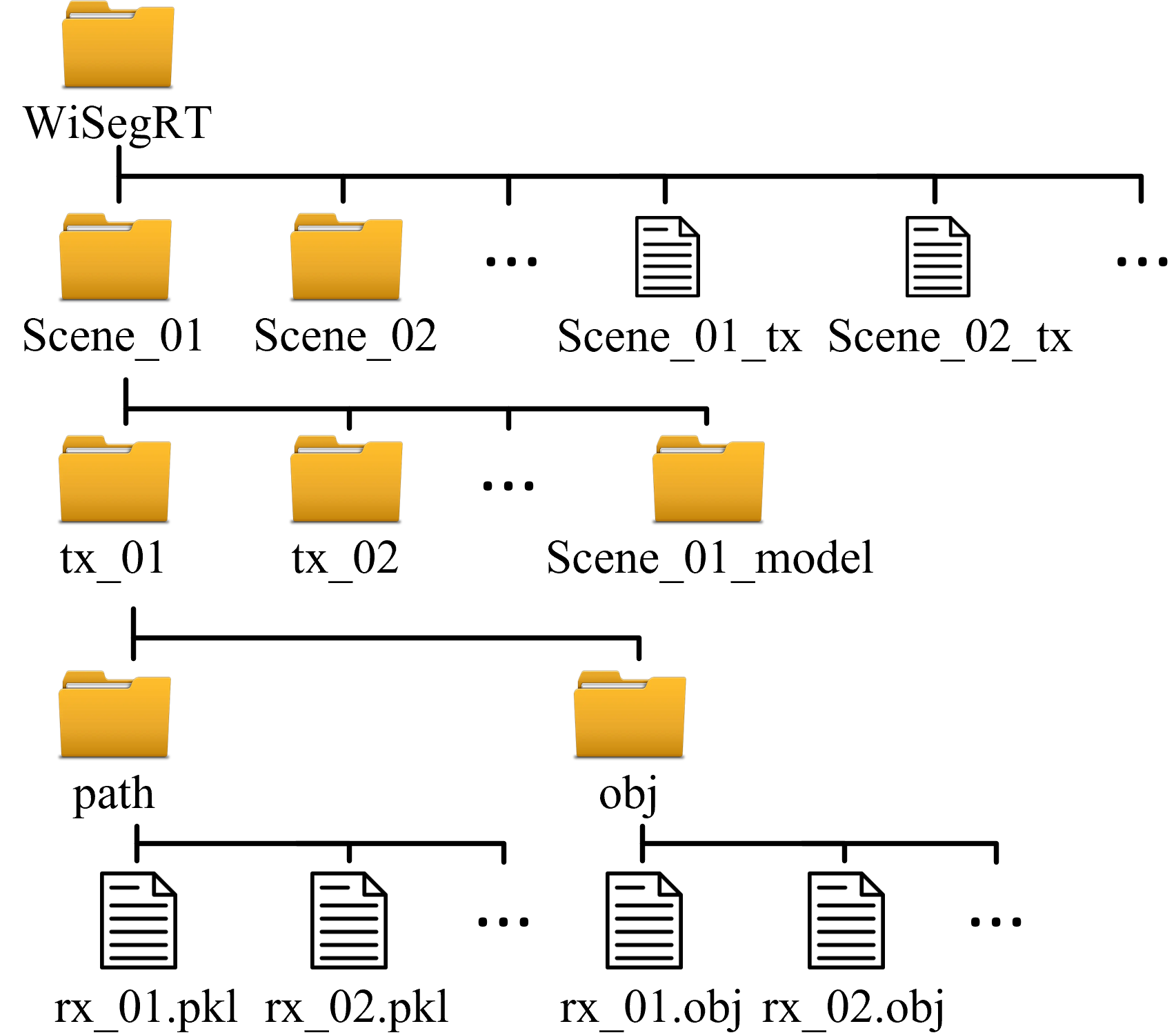}
    \caption{WiSegRT dataset file structure}
    \label{fig:file}
\end{figure}

\par
\begin{figure*}[ht]
    \centering
    \includegraphics[width=0.9\linewidth]{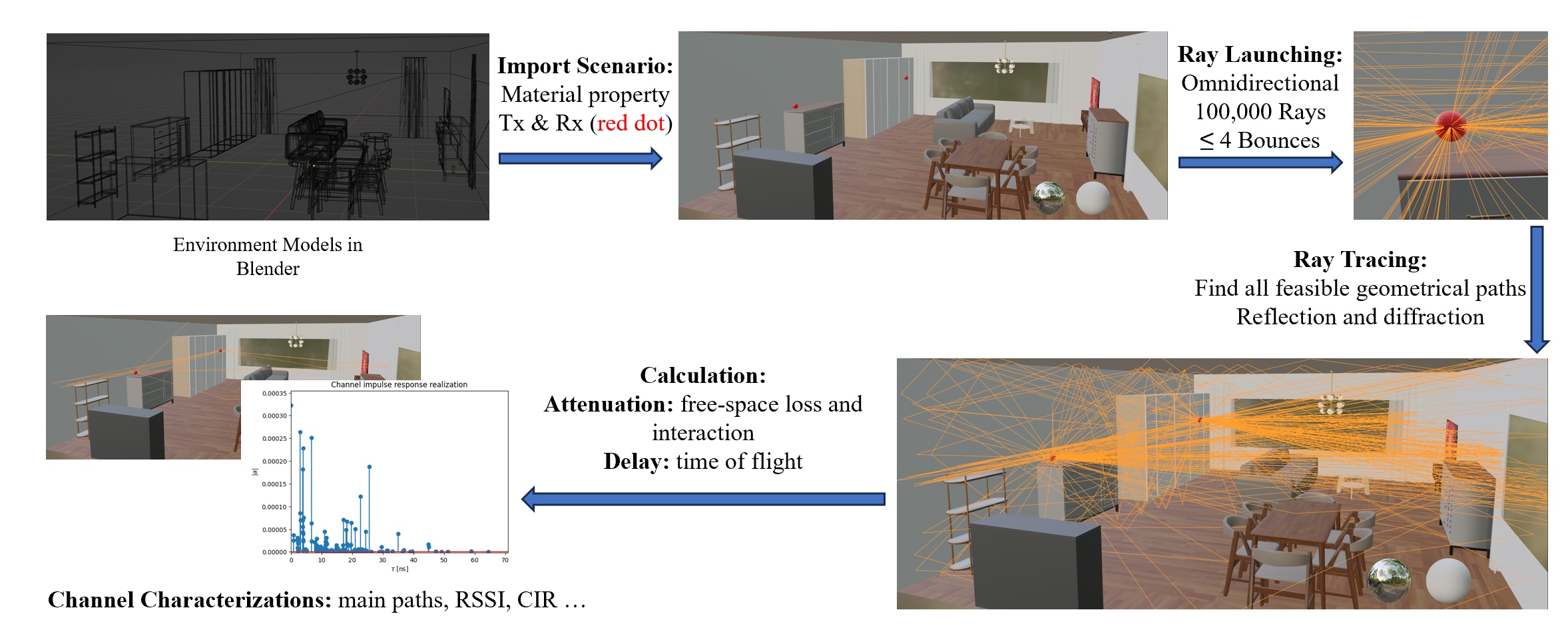}
    \caption{Flow of WiSegRT dataset generation }
    \label{fig:sysmodel}
\end{figure*}

 The structure of the WiSegRT dataset is illustrated in Fig. \ref{fig:file}. This dataset comprises ten distinct 3D scenes, each corresponding to a scene directory named as \texttt{\{Scene\_01, Scene\_02...\}}. For each scene, we set many reasonable Tx locations, which are recorded in the files, named as \texttt{\{Scene\_01\_tx, Scene\_02\_tx...\}}, under the main directory. We also create directories to store data for each Tx under each scene directory, named as \texttt{\{tx\_01,tx\_02...\}}. Rx locations are evenly distributed in the space, represented by 3D coordinates \texttt{[x,y,z]} with a 20cm resolution, i.e., \texttt{x = \{0m,0.2m,0.4m...\}}, \texttt{y = \{0m,0.2m,0.4m...\}}, \texttt{z = \{0m,0.2m,0.4m...\}}. We also provide 3D models for each scene, as \texttt{Scene\_01} contained in \texttt{Scene\_01\_model}. Each Tx directory is further divided into two sub-directories. The first one, \texttt{path}, contains the \texttt{.pkl} files which consist of the Tx position, Rx position, attenuation, delay, AoA and AoD for each Tx-Rx pair. The second sub-directory, \texttt{obj}, houses the \texttt{.obj} 3D model files of the geometrical paths between each Tx-Rx pair. These files can be loaded into renders for visualization, as demonstrated in Figure \ref{fig:sysmodel}. For instance, in \texttt{Scene\_01}, there are 30 transmitter positions and 3,325 receiver positions, resulting in a total of 99,750 communication pairs.

\section{Technical Details}
\par In this section, we give more technical details about the WiSegRT dataset. The system parameters are outlined in Table \uppercase\expandafter{\romannumeral1}, and the synthetic data generation flow is shown in Fig. \ref{fig:sysmodel}.

\begin{table}[h]
    \centering
    \textbf{Table \uppercase\expandafter{\romannumeral1}}~~System parameters of WiSegRT\\
    \setlength{\tabcolsep}{7mm}
    \begin{tabular}{cc}
        \hline
        \textbf{Parameter} & \textbf{Value} \\
        \hline
        Transmitter Antenna Pattern& Dipole\\
        Transmitter Antenna Polarization & Vertical \\
        Receiver Antenna Pattern & Isotropic\\
        Receiver Antenna Polarization& Cross\\
        Radio Frequency & 2.4 GHz\\
        Maximum Interaction & 4 \\
        Num. of Ray Launched & 100,000\\
        \hline
    \end{tabular}
    \label{tab:System_parameters}
\end{table}

\subsection{Radio Propagation and Ray Tracing}

\par 
Similar to how light radiates from a bulb and is perceived by the human eye, the radio are emitted from the Tx, propagate in the free space, interact with the objects, bounce zero or several times and reach the Rx. This process results in the radio propagating along various paths with distinct delays, attenuation and phase shifts, collectively known as the multi-path effect. The receiver must effectively manage this effect to reconstruct the transmitted information. 

\par As we build the environment models in \emph{Blender} and import the scenario to \emph{Sionna} with electromagnetic (EM) properties assigned to each object, together with the Tx and Rx positions, we can initiate the ray launching and tracing processes. To identify all feasible geometrical paths in the multipath-rich environments, we configured \emph{Sionna} to emit 100,000 rays omnidirectionally from the Tx. We set a maximum limit of 4 bounces, meaning that a ray is discarded if it does not reach the Rx after interacting with objects within 4 interactions.

\par After being emitted from the Tx, the ray will travel along straight lines in free space and obey the laws of geometrical optics when interacts with objects. In \emph{Sionna}, the ray has three types of interactions, e.g. reflection, diffraction, and scattering. Simple EM primer for understanding these interactions can be found from \emph{Sionna}'s document and textbooks. Generally, these interactions are highly related to the material EM properties (electric permittivity, magnetic permeability and conductivity), the incidence angle and the wave polarization. Surface roughness is another critical factor for scattering. By the Rayleigh criterion, if the height of surface irregularities is less than one-eighth of the wavelength, the surface can be considered smooth enough to neglect the scattering effects. In our scenarios, with a carrier frequency of 2.4 GHz corresponding to a wavelength of 0.125 m, most surfaces are deemed smooth, allowing us to disregard scattering path generation. However, for millimeter-wave and terahertz (THz) frequencies, surface roughness and scattering patterns must be considered and configured in \emph{Sionna}. Once all relevant factors are determined, \emph{Sionna} calculates the attenuation of the $N$ paths using TensorFlow, as shown in Eq. \eqref{attenuation}.
\begin{equation}
\begin{aligned} \label{attenuation}
\alpha_i = \sum\limits_{i=1}^N \frac{\lambda} {4\pi} C_{Rx}(\theta_{Rx,i}, \varphi_{Rx,i})^{H}T_i C_{Tx}(\theta_{Tx,i}, \varphi_{Tx,i}).
\end{aligned}
\end{equation}
\par The attenuation for each path in the radio propagation model is denoted by $\alpha_i$, where $i$ represents the index of the path. The antenna patterns of the Tx and Rx are represented by $C_{Tx}$ and $C_{Rx}$, respectively. These patterns are utilized in determining the gains of the antennas based on the AoA and AoD, which are denoted by $(\theta_{Rx,i}, \varphi_{Rx,i})$ and $(\theta_{Tx,i},\varphi_{Tx,i})$. The wavelength of the radio signal is represented by $\lambda$, and the transformations that occur between the Tx and Rx antennas are encapsulated within the matrix $T_i$. These transformations include changes in direction and amplitude attenuation due to free space propagation and interactions with objects. The delay $\tau_i$ for the $i$ th path is computed based on the total path length and the speed of light.

\par It is important to note that the latest \emph{Sionna} system does not support modeling of refraction. This limitation is a result of the challenges associated with accurately modeling the internal structure and heterogeneous materials of objects.  For large-scale outdoor scenes this is typically not a significant concern, as the objects consist mainly of large size buildings with glass curtain walls, dense concrete walls, and intricate inner structures. Radio waves generally cannot penetrate these objects or reflect from their inner surfaces \cite{mmsv_kamari2023mmsv}. In contrast, for indoor environments, the effects of the object's internal parts may become more pronounced due to the smaller sizes of objects. Further research is needed to develop accurate measurement and modeling for these interactions.

\subsection{Radio and Channels}
\par As we have the attenuation and delay of each path, we can then transfer them to the CIR as each path corresponding to an impulse, as shown in Eq. \eqref{CIR}.
\begin{equation}
\begin{aligned} \label{CIR}
h(\tau) = \sum\limits_{i=1}^N \alpha_i\delta(\tau-\tau_i).
\end{aligned}
\end{equation}
\par This is the CIR of the physical wireless channel, which can only be ``seen" by the radio frequency front-end. For wireless systems, the focus is more on the baseband equivalent channel and its impulse response, which can be written as Eq. \eqref{baseband-CIR}.
\begin{equation}
\begin{aligned} \label{baseband-CIR}
h_b(\tau) = \sum\limits_{i=1}^N \alpha_i e^{-j2\pi f\tau_i}\delta(\tau-\tau_i).
\end{aligned}
\end{equation}
\par Here, the $f$ refers to the carrier frequency. The main differences are the phase shifts on each path, as indicated by the exponential term $e^{-j2\pi f\tau_i}$. This means that for a given scenario where the paths, $\alpha_i$ and $\tau_i$ are determined, the phase shift caused by carrier frequency differences, like the differences among sub-carriers, may largely change the baseband equivalent channel, which is known as the frequency selective fading.

\section{Comparison of the WiSegRT with Different Settings}

\par As the WiSegRT dataset has very detailed models of objects and precise assignment of EM material properties, it introduces significant overhead in generating additional data and further tasks due to the large number of primitives the object models contain. This level of detail differs from most previous works, which often use vague and simple models and may only consider assigning material for large, strong reflective objects, such as window and metal refrigerator. 

\par In this section, we focus on the ``\texttt{Scene\_01}'' to compare the original WiSegRT dataset, denoted as ``ori" (original) scenario, with two other scenarios generated under different settings. To validate the necessity of fine 3D models, we constructed an empty version of the same room without any furniture or appliances, denoted as ``emp" (empty) scenario, which only contains the concrete wall, glass windows, and wood floor with corresponding EM properties, similar to \cite{3d_kocevska2023identification}. To assess the impact of the fineness of environment, we created a simplified version of the room that includes only the main furniture with less precise 3D model and material assignments, denoted as ``sim" (simplified) scenario. We designate the ``ori"  as the ground truth, which is the closest representation of reality. The other two scenarios, due to their partial omission of information, are expected to deviate more from actual conditions.  

\begin{figure}[h]
    \centering
    \includegraphics[width=0.95\linewidth]{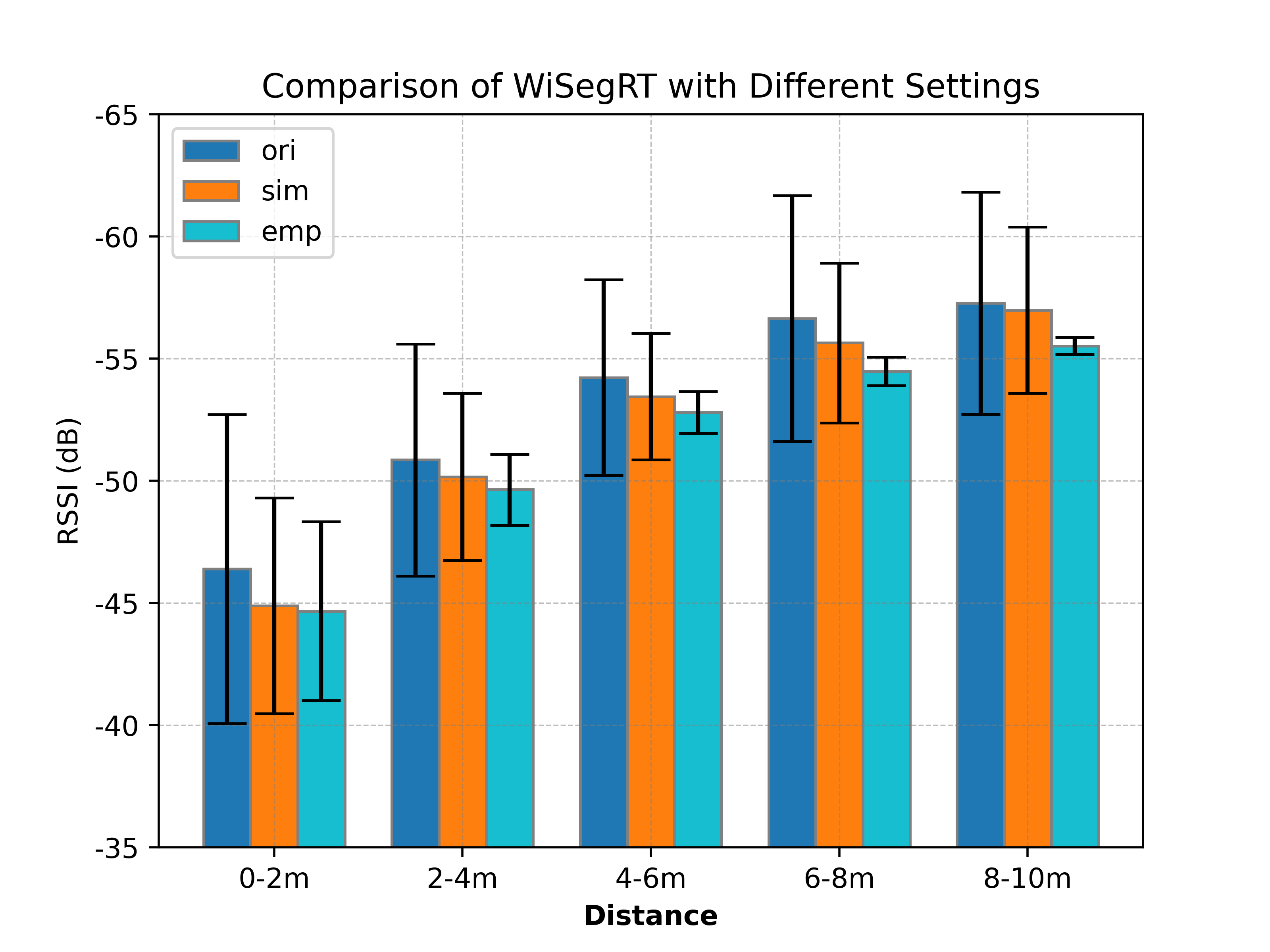}
    \caption{Comparison of total received power at different distances}
    \label{fig:rssi}
\end{figure}

\par Fig. \ref{fig:rssi} presents a comparison of the average RSSI in different Tx-Rx distance ranges. The RSSI of each Tx-Rx pair is calculated from the total received power of all paths. It is evident from the figure that while the average RSSI of the three scenarios does not significantly differ on the decibel (dB) scale, there is a notable difference in their standard deviation (SD). As the Tx-Rx pairs are grouped by their distance, the SD of RSSI is predominantly attributed to the environmental multi-path effects and the blockage of the line-of-sight. This SD serves as an indicator of how accurately each scenario represents the real-world radio propagation characteristics.

\par In the ``sim" scenario, we can find a reduction in SD compared to the ground truth, which can be attributed to the absence of objects and less detailed material segmentation. Besides, the average RSSI values on dB scale also differ a little between the ``ori" and ``sim" scenarios, suggesting that fine-grained 3D objects and their material EM properties have a considerable impact on RSSI distribution. In the ``emp" scenario, the degradation is worse. As we remove all the inner objects, the simulated RSSI values exhibit a significantly lower SD at distances beyond 2m, which means this kind of ``emp" setting fails to accurately reflect the RSSI distribution in real-world.



\section{Possible Applications}

\subsection{Vision-based Radio Tracing}
\par Ray tracing methods, originally designed for graphic rendering to construct digital images, can also be applied to radio propagation modeling. However, significant differences exist between graphical rendering and radio simulations. 

\par  In terms of EM waves, graphic rendering mainly deals with visible light, which spans a broad frequency band (380 -- 750 THz) and contains a complex mixture of wavelengths. Despite the vast spectrum processed by the human eye, our vision has relatively low temporal and spectral resolution, capturing only 24-bit color information (8 bits each for red, green, and blue) and processing images at roughly 24 frames per second. Conversely, radio frequencies operate within a comparatively ``narrow" band and are discrete, requiring a higher resolution in both frequency and time domains, such as 15 kHz and 50 $\mu s$ in LTE. This contrast highlights that graphical rendering is more concerned with the statistical characteristics of each path, whereas radio simulations prioritize accurate attenuation, delay, and phase shift of each path.

\par On the ``receiver" side, graphic rendering requires superior angular resolution and millions of rays to render a high-resolution image. But for radio, the receiver focuses more on how the wireless channel distorts the transmitted EM wave which can be represented by the CIR. In normal indoor scenarios, the number of geometrical paths, also the number of impulses of the CIR, rarely exceeds 200. The number of main paths, those with the least attenuation, rarely exceeds 20. Due to these disparities, there is a clear need to develop radio tracing methodologies that can efficiently identify main paths instead of emitting an excessive number of rays and subsequently discarding most of them.

\par
\begin{figure}[h]
    \centering
    \includegraphics[width=1.0\linewidth]{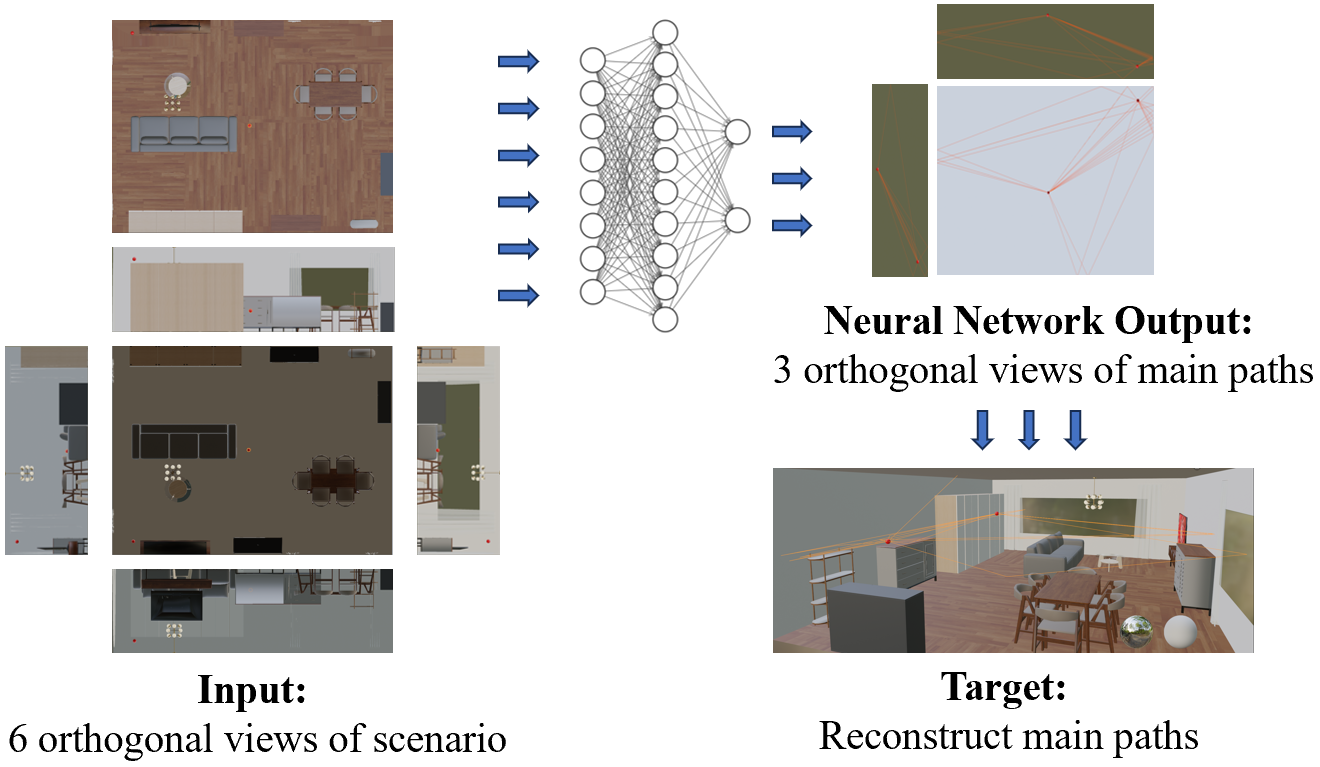}
    \caption{Radio tracing with vision data and ML}
    \label{fig:radio_tracing}
\end{figure}

 Neural networks have demonstrated the ability to mimic the radio propagation pattern in 2D environment, as indicated in \cite{2d_radio_levie2021radiounet}. It is plausible that neural networks can also learn to identify main paths in the orthogonal views of 3D environments. As shown in Fig. \ref{fig:radio_tracing}, our preliminary scheme takes the orthogonal images from 6 directions \texttt{(+x,-x,+y,-y,+z,-z)} as the representation of the environment, along with the locations of the Tx and Rx. The output images are orthogonal images of the path objects from 3 directions \texttt{(+x,+y,+z)}, which are adequate for path recovery. Since the rendered images contain information about object structure and material, it is possible for neural network to implicitly learn the correlation between environmental visuals and main paths.However, this concept is still in its early stages, and the outcomes thus far are constrained by both theoretical and implementation factors in machine learning.
\subsection{Indoor Radio Sensing}
\par As EM wave has been deployed in radars for sensing from 1930s, employing radio for indoor sensing is a natural progression, driven by the widespread adoption of wireless devices. Many pioneering studies  have leveraged radio waves for indoor sensing applications, such as monitoring human pose and even vital signs. As we provide the 3D model of the environment, it supports further scenario customization, including the addition of human figures, editing their actions, and even varying levels of perspiration as it largely changes the EM properties of skin and clothing. By understanding how radio waves interact with target and subsequently reach the Rx, we can build more physics-informed approaches rather than relying solely on end-to-end models from the noisy measurements.

\section{Conclusions}
In this paper, we introduced the WiSegRT, a precise indoor radio propagation dataset. Compared with the simple settings deployed by previous works to generate dataset, the WiSegRT dataset provides an enhanced representation of the complex characteristics of real-world radio propagation. With the virtual-reality-like scene models, WiSegRT supports various ML-based wireless tasks, such as the vision-based radio tracing and ray-level indoor radio-sensing. Future works will focus on developing model editing libraries to support digital-twin-styled tasks.

\bibliographystyle{IEEEtran}
\bibliography{lib}

\end{document}